# Use of the journal impact factor for assessing individual articles: Statistically flawed or not?

Ludo Waltman and Vincent A. Traag

Centre for Science and Technology Studies, Leiden University, The Netherlands
{waltmanlr, v.a.traag}@cwts.leidenuniv.nl

Most scientometricians reject the use of the journal impact factor for assessing individual articles and their authors. The well-known San Francisco Declaration on Research Assessment also strongly objects against this way of using the impact factor. Arguments against the use of the impact factor at the level of individual articles are often based on statistical considerations. The skewness of journal citation distributions typically plays a central role in these arguments. We present a theoretical analysis of statistical arguments against the use of the impact factor at the level of individual articles. Our analysis shows that these arguments do not support the conclusion that the impact factor should not be used for assessing individual articles. Using computer simulations, we demonstrate that under certain conditions the number of citations an article has received is a more accurate indicator of the value of the article than the impact factor. However, under other conditions, the impact factor is a more accurate indicator. It is important to critically discuss the dominant role of the impact factor in research evaluations, but the discussion should not be based on misplaced statistical arguments. Instead, the primary focus should be on the socio-technical implications of the use of the impact factor.

## 1. Introduction

The journal impact factor (IF) is the most commonly used indicator for assessing scientific journals. IFs are calculated based on the Web of Science database. They are reported each year in the Journal Citation Reports published by Clarivate Analytics. Essentially, for a certain year $y$, the IF of a journal equals the average number of citations received in year $y$ by articles published in the journal in years $y - 1$ and $y - 2$. Although the IF is an indicator at the level of journals, it is used not only for assessing journals as a whole, but also for assessing individual articles in a journal.

IF-based assessments of individual articles are usually used to evaluate the researchers or the institutions by which the articles have been authored.

There is a lot of criticism on the IF and its use in research evaluations (e.g., DORA, 2013; Seglen, 1997; Vanclay, 2012). One of the most common concerns relates to the use of the IF for assessing individual articles. It is often argued that from a statistical point of view it is incorrect, or at least highly problematic, to use journal-level indicators such as the IF in the assessment of individual articles (e.g., Garfield, 2006; Gingras, 2016; Larivière et al., 2016; Leydesdorff, Wouters, & Bornmann, 2016; Seglen, 1992, 1997; Zhang, Rousseau, & Sivertsen, 2017). This point is also made in the well-known San Francisco Declaration on Research Assessment (DORA, 2013). The argument is that the IF is a journal-level indicator and that it therefore tells us something about a journal as a whole, but not about an individual article in a journal. Typically the argument is supported by pointing out that the distribution of citations over the articles in a journal is highly skewed, with a small share of the articles in a journal receiving a large share of the citations (Seglen, 1992, 1997). The IF of a journal therefore is not representative of the number of citations of an individual article in the journal. According to Curry (2012), a prominent critic of the IF, the use of the IF at the level of individual articles reflects 'statistical illiteracy'.

In this paper, we analyze in detail the above statistical argument against the use of the IF for assessing individual articles. We point out that the argument does not logically lead to the conclusion that the IF should not be used at the level of individual articles. The argument leads to this conclusion only when additional assumptions are made. If one considers these assumptions to be reasonable, the use of the IF for assessing individual articles should indeed be rejected. However, if one makes alternative assumptions, the opposite conclusion may be reached and the use of the IF for assessing individual articles may be argued to be preferable over the use of indicators defined at the level of individual articles, such as the number of citations of an article.

The aim of this paper is not to argue in favor of or against either the IF in general or the specific use of the IF for assessing individual articles. The analysis that we present does not enable us to draw a general conclusion on the appropriateness of IF-based assessment of individual articles. Rather, our aim is to criticize the statistical objections typically raised against the use of the IF at the level of individual articles. We argue that these objections are misguided. In our view, it is important to critically discuss the dominant role of the IF in research evaluations, but the discussion should not be based on the misplaced argument that the use of a journal-level indicator for assessing individual articles is statistically flawed. Instead, the discussion should for instance consider the calculation of the IF (e.g., consistency of the numerator and denominator and choice of publication and citation windows), its transparency, and its vulnerability to manipulation. Even more importantly, the discussion should focus on the socio-technical implications of the pervasive use of the IF.

Although the discussion in this paper focuses on the IF, we emphasize that the discussion also applies to other citation-based indicators for journals. Indicators such as Eigenfactor Score and Article Influence Score (West, Bergstrom, & Bergstrom, 2010), Source Normalized Impact per Paper (SNIP; Moed, 2010; Waltman, Van Eck, Van Leeuwen, & Visser, 2013), Scimago Journal Rank (SJR; González-Pereira, Guerrero-Bote, & Moya-Anegón, 2010; Guerrero-Bote & Moya-Anegón, 2012), geometric IF (Thelwall & Fairclough, 2015), and CiteScore (James, Colledge, Meester, Azoulay, & Plume, 2019) differ from the IF in various ways. However, like the IF, these indicators are all defined at the level of journals. The discussion on the use of journal-level indicators at the level of individual articles is therefore equally relevant for these indicators as it is for the IF. Our focus in this paper is on the IF simply because the IF is the most commonly used journal-level indicator, and consequently also the indicator that is debated most heavily. We refer to Waltman (2016, Section 8) for an overview of the literature on citation-based indicators for journals.

This paper is organized as follows. Section 2 gives an overview of the discussion on the use of the IF for assessing individual articles. Section 3 provides an illustrative example analyzing the use of the IF at the level of individual articles. This is followed in Section 4 by a more general conceptual discussion on the use of the IF for assessing individual articles. The illustrative example in Section 3 and the conceptual discussion in Section 4 aim to make clear that from a statistical point of view the use of the IF at the level of individual articles does not need to be wrong. Section 5 presents computer simulations to further illustrate this point. Finally, in Section 6, we discuss our findings and summarize our conclusions.

## 2. Background

There is a sizeable literature discussing the IF and its use in research evaluations (for a recent overview, see Larivière & Sugimoto, 2019). The discussion has partly focused on technical and statistical issues in the calculation of the IF (e.g., Glänzel & Moed, 2002; Seglen, 1997), such as the definition of so-called 'citable items' in the denominator of the IF (e.g., Moed & Van Leeuwen, 1995, 1996) and the time window based on which the IF is calculated (e.g., Glänzel & Schoepflin, 1995; Moed, Van Leeuwen, & Reedijk, 1998). In addition, there has also been discussion about the transparency of the IF (e.g., PLoS Medicine Editors, 2006; Rossner, Van Epps, & Hill, 2007; Vanclay, 2012), the vulnerability of the IF to manipulation (e.g., Chorus & Waltman, 2016; Martin, 2016; Wilhite & Fong, 2012), and the dominant role of the IF in research evaluations (e.g., McKiernan, Schimanski, Nieves, Matthias, Niles, & Alperin, 2019; Quan, Chen, & Shu, 2017; Rushforth & De Rijcke, 2015). An extensive discussion about the IF has taken place in a special issue of *Scientometrics* (Braun, 2012). This discussion was triggered by a critical paper about the IF by Vanclay (2012). The producers of the IF have also repeatedly contributed to discussions about the IF (e.g., Garfield, 1996, 2006; Pendlebury, 2009; Pendlebury & Adams, 2012; Wouters et al., 2019).

In this paper, we restrict our attention to statistical objections against the use of the IF for assessing individual articles. Below we first review some literature that argues against IF-based assessment of individual articles. We then discuss a few sources which suggest that there are some limited opportunities for IF-based assessment of individual articles.

### 2.1. Statistical objections against the use of the impact factor for assessing individual articles

Statistical objections against IF-based assessment of individual articles go back at least to classical papers by Seglen (1992, 1997). Seglen shows that the distribution of citations over the articles in a journal is highly skewed. He then draws the following conclusion (Seglen, 1992, p. 631):

> The great variability in citedness within a journal has important implications for the significance attached to the journal impact factor. In several countries, this easily available factor has been used in academic evaluations of individual

> scientists, on the implicit premise that the impact factor of the journal is representative of its constituent articles, and hence, of the article authors. The skewness of the journal article distributions shows that this premise does not hold true: only a minor fraction of the articles are cited anywhere near the journal mean ... Assigning the same value to all articles in a journal will overestimate the less influential and underestimate the more influential articles, thus effectively leveling out the very differences that evaluation procedures should seek to identify.

Eugene Garfield, who created the IF in the early days of the Science Citation Index, draws a similar conclusion (Garfield, 2006, p. 92):

> Typically, when the author's work is examined, the impact factors of the journals involved are substituted for the actual citation count. Thus, the journal impact factor is used to estimate the expected count of individual papers, which is rather dubious considering the known skewness observed for most journals.

In 2013, the San Francisco Declaration on Research Assessment (DORA) was published. It has attracted a lot of attention and support. DORA strongly rejects the use of the IF for assessing individual articles. A number of arguments are given, one of them being that "citation distributions within journals are highly skewed", leading to the recommendation not to "use journal-based metrics, such as Journal Impact Factors, as a surrogate measure of the quality of individual research articles (or) to assess an individual scientist's contributions" (DORA, 2013). DORA also recommends journal publishers to "make available a range of article-level metrics to encourage a shift toward assessment based on the scientific content of an article rather than publication metrics of the journal in which it was published".

In line with DORA, in his monograph on bibliometrics and research evaluation, Gingras (2016) also uses the skewness of citation distributions to argue against the use of the IF for assessing individual articles (p. 47–48):

> The IF remains a measure related to the journal, not to the articles it contains. The fundamental reason that makes it a flawed indicator of the value of

> individual articles is that the distribution of actual citations received by the articles published in a given journal follows a power law similar to that of Alfred Lotka for productivity, which means that most articles are in fact cited very little. Only a few are very highly cited, and they inflate the value of the IF ... If one wants to measure the quality or visibility of a particular item, one must look at the citations actually received in the years following its publication. But that of course takes time, and those who prefer 'quick and dirty' evaluation do not want to wait three to five years. So they use the IF of the journal in which the papers are published as a proxy of their quality and impact, even though such a measure is totally inappropriate.

A high-profile paper by Larivière et al. (2016) again draws attention to the skewness of the distribution of citations over the articles in a journal. The authors recommend that, when a journal publishes its IF, it should also publish the underlying citation distribution. In this way, awareness will be drawn to the skewness of the citation distribution, and this skewness can then be taken into account in the interpretation of the IF. Like the sources discussed above, Larivière et al. regard the skewness of the citation distribution of a journal as an argument against the use of the IF for assessing individual articles. They observe that "for all journals there are large numbers of papers with few citations and relatively few papers with many citations", which they argue "underscores the need to examine each paper on its own merits and serves as a caution against over-simplistic interpretations of the JIF" (p. 5).

In a recent paper by Zhang et al. (2017), the work by Seglen (1992, 1997) is revisited. Seglen's empirical findings are confirmed based on a much larger amount of data, leading to the following conclusion (p. 14–15):

> Although some journals are certainly more prestigious, attractive and selective than others, one should not infer the quality of the individual article from the status of the journal. Moreover, even if citations are taken as an indication of quality, the citation impact of a journal remains a weak predictor of the citation impact of each of its articles. Consequently, individual contributions should not be evaluated by where they are published.

Leydesdorff et al. (2016) also present a statistical objection against the use of the IF for assessing individual articles. Their objection does not relate to the skewness of citation distributions. Instead, it is based on the concept of ecological fallacy (p. 2140):

> The use of the JIF for the evaluation of individual papers provides an example of the so-called "ecological fallacy" ...: inferences about the nature of single records (here: papers) are deduced from inferences about the group to which these records belong (here: the journals where the papers were published). However, an individual child can be weak in math in a school class which is the best in a school district. Citizen bibliometricians ... may nevertheless wish to continue to use the JIF in research evaluations for pragmatic reasons, but this practice is ill-advised from the technical perspective of professional bibliometrics.

Paulus, Cruz, and Krach (2018) also use the concept of ecological fallacy to criticize the use of the IF at the level of individual articles.

### 2.2. Limited opportunities for the use of the impact factor for assessing individual articles

Using statistical arguments similar to the ones presented above, most scientometricians reject the use of the IF for assessing individual articles. However, some scientometricians argue that there is some room for assessing individual articles using the IF or some other journal-level indicator.

According to Abramo, D'Angelo, and Di Costa (2010, p. 832), "there is an agreement among scholars on the superiority of citations over impact factor as proxy of quality of publications for 'old' articles". However, for recent articles, Abramo et al. argue that the situation is different:

> Citations observed at a moment too close to the date of publication will not necessarily offer a proxy of quality that is preferable to impact factor. Yet bibliometric evaluation exercises ... should be based on observations of the most recent possible past. For evaluations over periods that are very close in time to the date of conducting the exercise, and especially in certain

> disciplines, the impact factor can thus be a predictor of the real impact of an article, and possibly a better one than citations.

A similar argument is made by Levitt and Thelwall (2011). Rather than choosing between the number of citations of an article and the IF of the journal in which an article has appeared, Levitt and Thelwall suggest to combine the number of citations and the IF into a hybrid indicator. In the context of providing indicators to peer review panels in the UK Research Excellence Framework, Levitt and Thelwall reach the following conclusion (p. 307):

> Particularly for very recently published articles, an indicator based on the average of the standard indicator of citation and the IF of the journal ... could form the basis of a useful indicator for peer review panels.

Ancaiani et al. (2015) discuss how the Italian research evaluation exercise takes into account both the number of citations of an article and the IF of the journal in which an article has appeared. In line with the ideas of Abramo et al. (2010) and Levitt and Thelwall (2011), the IF plays a prominent role especially in the assessment of recent articles. When the number of citations and the IF provide conflicting information, the IF is given more weight in the case of recent articles, while the number of citations has more weight in the case of older articles.

Another perspective is provided by Moed (2005) in his monograph on citation analysis and research evaluation. According to Moed, assessing articles using journal-level indicators is acceptable, but the assessment should focus on the entire oeuvre of a research group rather than on individual articles. Moreover, Moed emphasizes that journal-level indicators reflect a different aspect of the performance of a research group than article-level indicators (p. 84–85):

> Journal impact is a performance aspect in its own right, but cannot be used to predict actual citation rates. The extent to which groups of scientists publish their output in the more prestigious, or even the 'top' journals in their fields, is often viewed as an important aspect of scientific research performance. (An) indicator of the impact of a group's journal packet ... can be validly used to assess this aspect.

## 3. Illustrative example

In this section, we present a simple illustrative example analyzing the use of the IF for assessing individual articles. The example introduces some of the key ideas that will play an important role in the conceptual discussion in Section 4 and in the computer simulations in Section 5. Before presenting the example, we first discuss the difference between observable and non-observable concepts in citation analysis.

### 3.1. Observable and non-observable concepts

In order to have a careful and precise discussion on the use of the IF for assessing individual articles, it is essential to distinguish between observable and non-observable concepts in citation analysis (for a similar idea in a somewhat different context, see Waltman, Van Eck, & Wouters, 2013). Important observable concepts are the number of citations of an article and the IF of a journal. These observable concepts are important not so much because they are of interest themselves, but mainly because they may tell us something about certain non-observable concepts that we are interested in. In the context of the assessment of scientific articles, examples of these non-observable concepts could be the quality, the impact, and the influence of an article. The general idea of citation analysis is that an observable concept, such as the number of citations of an article, provides an approximate representation of a non-observable concept, such as the impact of an article. The observable concept is then regarded as an indicator of the non-observable concept. The number of citations of an article for instance is often regarded as an indicator of the impact of the article. Likewise, the IF of a journal is sometimes seen as an indicator of the quality of the journal.

The use of a certain observable concept as an indicator of a certain non-observable concept often causes debate. There usually is disagreement on whether the observable concept provides a sufficiently close approximation of the non-observable concept. For instance, some may consider the number of citations of an article to be a suitable indicator of the impact of the article, but others may disagree and may argue that citations do not provide a sufficiently close approximation of impact. At a more fundamental level, the difficulty is that non-observable concepts typically lack a clear and unambiguous definition. The concepts of quality, impact, and influence for instance are understood differently by different people, making it challenging to agree on the use of citations as an indicator of any of these concepts.

In this paper, we do not want to enter the debate about which non-observable concepts may or may not be represented by the number of citations of an article. Instead, we start from the idea that assessing an article is equivalent to determining the value of the article, where we use value as a general non-observable concept that, depending on the precise criterion based on which one wants articles to be assessed, may for instance be understood as quality, impact, influence, importance, or usefulness. The main point that we want to make in this paper does not depend on the specific understanding that one has of the concept of value, and therefore there is no need to provide a precise definition of this concept. However, in Subsection 4.2, we will say a bit more about the implications of different ways in which the concept of value can be understood.

For further discussion on the conceptual foundation of citation analysis, we refer to Bornmann and Daniel (2008), De Bellis (2009, Chapter 7), Moed (2005, Chapters 15–17), Nicolaisen (2007), and Tahamtan and Bornmann (2019).

### 3.2. Example

We now provide a simple example comparing the assessment of articles based on either the IF of the journal in which they have appeared or the number of citations they have received. The example introduces some of the key ideas that will be further elaborated in Sections 4 and 5. It also illustrates the importance of making a careful distinction between observable and non-observable concepts.

The situation that we analyze in our example is an extreme simplification of reality (for a somewhat similar type of analysis, see Waltman et al., 2013). Some may regard this as a weakness of the example. However, we regard it as a strength, because the extreme simplification enables us to focus on the most essential issues, without being distracted by irrelevant details.

We consider a situation in which the value of an article is either low or high and in which an article is either lowly cited or highly cited. There are 200 articles. Of these articles, 100 are of low value and 100 are of high value. Likewise, 100 are lowly cited and 100 are highly cited. Furthermore, there are just two journals, journal A and journal B. Each journal has published 100 articles.

Our aim is to identify as accurately as possible the articles that are of high value. As pointed out in Subsection 3.1, the value of an article is a non-observable concept. This means that high-value articles cannot be directly identified. We therefore compare two approaches that try to identify these articles in an indirect way. One approach is to select all articles that are highly cited. The other approach is to select all articles that have appeared in the journal with the higher IF. In the situation in which each article is either lowly cited or highly cited, the journal with the higher IF is the journal with the larger share of highly cited articles. We want to find out which of the above two approaches for identifying high-value articles is more accurate. For each approach, we quantify the accuracy of the approach by determining the probability that a selected article is indeed of high value.

The number of citations of an article may provide an approximate representation of the value of an article. Because the representation is approximate, being highly cited does not need to coincide with being of high value. In the first scenario that we consider (i.e., scenario 1), 90% of the articles that are of high value are highly cited. The other 10% are lowly cited. Conversely, 90% of the articles that are of low value are lowly cited. The other 10% are highly cited. This information is summarized in Table 1.

Suppose that 80 articles in journal A are of high value, while only 20 articles in journal B are of high value. The remaining articles in both journals are of low value. This yields the situation presented in Table 2. As can be seen in the table, the number of highly cited articles in journal A equals $90\% \times 80 + 10\% \times 20 = 74$. On the other hand, journal B has published $90\% \times 20 + 10\% \times 80 = 26$ highly cited articles. Consequently, journal A has published a larger share of highly cited articles than journal B, and therefore journal A has a higher IF than journal B.

If we choose to identify high-value articles based on the IF, we select all 100 articles in journal A, which yields 80 high-value articles. The other approach is to identify high-value articles based on an article's number of citations. If we choose this approach, we select all 100 highly cited articles. 90% of these articles are of high value, so this results in 90 high-value articles. Hence, in scenario 1, it is more accurate to identify high-value articles based on an article's number of citations than based on the IF. This is in agreement with commonly used statistical arguments against the use of the IF for assessing individual articles.

Table 1. Probability that an article is lowly or highly cited conditional on the article being of low or high value (scenario 1).

| | Lowly cited | Highly cited |
|---|---|---|
| Low value | 0.9 | 0.1 |
| High value | 0.1 | 0.9 |

Table 2. Breakdown of the number of articles in journals A and B by value and number of citations (scenario 1).

Journal A

| | Lowly cited | Highly cited | Total |
|---|---|---|---|
| Low value | 18 | 2 | 20 |
| High value | 8 | 72 | 80 |
| Total | 26 | 74 | 100 |

Journal B

| | Lowly cited | Highly cited | Total |
|---|---|---|---|
| Low value | 72 | 8 | 80 |
| High value | 2 | 18 | 20 |
| Total | 74 | 26 | 100 |

We now consider a second scenario (i.e., scenario 2). In this scenario, instead of 90% only 70% of the high-value articles are highly cited. The other 30% are lowly cited. Of the low-value articles, 70% are lowly cited and 30% are highly cited. Like in scenario 1, 80 articles in journal A are of high value, while only 20 articles in journal B are of high value. All other articles are of low value. Scenario 2 is summarized in Tables 3 and 4.

To what extent does scenario 2 lead to different outcomes than scenario 1? In scenario 2, journals A and B have published respectively $70\% \times 80 + 30\% \times 20 = 62$ and $70\% \times 20 + 30\% \times 80 = 38$ highly cited articles. Hence, like in scenario 1, journal A has a higher IF than journal B. If we choose to identify high-value articles based on the IF, we select all 100 articles in journal A. This yields 80 high-value articles, which is identical to the outcome obtained in scenario 1. On the other hand, if we choose to identify high-value articles based on an article's number of citations, we select all 100 highly cited articles. In scenario 2, only 70% of these articles are of high value, and therefore we obtain only 70 high-value articles.

Table 3. Probability that an article is lowly or highly cited conditional on the article being of low or high value (scenario 2).

| | Lowly cited | Highly cited |
|---|---|---|
| Low value | 0.7 | 0.3 |
| High value | 0.3 | 0.7 |

Table 4. Breakdown of the number of articles in journals A and B by value and number of citations (scenario 2).

Journal A

| | Lowly cited | Highly cited | Total |
|---|---|---|---|
| Low value | 14 | 6 | 20 |
| High value | 24 | 56 | 80 |
| Total | 38 | 62 | 100 |

Journal B

| | Lowly cited | Highly cited | Total |
|---|---|---|---|
| Low value | 56 | 24 | 80 |
| High value | 6 | 14 | 20 |
| Total | 62 | 38 | 100 |

Importantly, the conclusion that we reach in scenario 2 is the opposite of the conclusion drawn in scenario 1. In scenario 2, identifying high-value articles based on an article's number of citations is less accurate than identifying high-value articles based on the IF. The accuracy of citations as an indicator of the value of an article is lower in scenario 2 than in scenario 1, but this difference in the accuracy of citations does not affect the accuracy of the IF. This explains why the two scenarios yield opposite conclusions and why in scenario 2 the IF is a more accurate indicator of the value of an article than the number of citations of the article.

Of course, the situation analyzed in the above example is an extreme simplification of reality. Nevertheless, the example shows that the number of citations of an article is not necessarily a more accurate indicator of the value of the article than the IF of the journal in which the article has appeared. Which of the two indicators is more accurate depends on the degree to which citations provide an accurate representation of the value of an article. In the next two sections, we will study this in more detail, first by providing a conceptual discussion and then by presenting computer simulations.

## 4. Conceptual discussion

In the previous section, we provided an illustrative example of a situation in which it is possible that the IF of the journal in which an article has appeared is a more accurate indicator of the value of the article than the number of citations of the article. The situation analyzed in the example in the previous section is an extreme simplification of reality. As we have seen in Section 2, in discussions on the use of the IF for assessing individual articles, the skewness of the distribution of citations over the articles in a journal usually plays a crucial role. The skewness of journal citation distributions was not taken into account in the simple example presented in the previous section. In this section, we provide a more general conceptual discussion on the use of the IF for assessing individual articles. The skewness of journal citation distributions is a key element in this discussion.[1]

### 4.1. Two scenarios

Like in the example presented in the previous section, the distinction between the value of an article and the number of citations of an article is essential. We again consider two scenarios. In scenario 1, the number of citations of an article is a more accurate indicator of the value of the article than the IF of the journal in which the article has appeared. Scenario 2 represents the opposite situation. In both scenarios, journal citation distributions are highly skewed.

Scenario 1 can be summarized in the following three points:

1. The number of citations of an article is a *relatively accurate* indicator of the value of the article.

[1] Our use of the term 'skewness' in this paper follows the literature discussed in Subsection 2.1. However, we note that it would actually be more appropriate to consider the variance rather than the skewness of journal citation distributions. If the citation distribution of a journal is perfectly symmetrical (and therefore completely non-skewed) but has a high variance, the IF would still not be representative of the number of citations of an individual article in the journal. Presumably, many scientometricians would then still have statistical objections against the use of the IF at the level of individual articles.

2. Journals are *rather heterogeneous* in terms of the values of the articles they publish.
3. The skewness of journal citation distributions results mainly from point 2.

Compared with scenario 1, scenario 2 offers an opposite explanation of the skewness of journal citation distributions:

1. The number of citations of an article is a *relatively inaccurate* indicator of the value of the article.
2. Journals are *fairly homogeneous* in terms of the values of the articles they publish.
3. The skewness of journal citation distributions results mainly from point 1.

In scenario 1, the number of citations of an article and the value of an article are strongly correlated. The skewness of a journal citation distribution therefore reflects the skewness of the distribution of the values of the articles in a journal. The IF is not representative of the number of citations of an individual article in a journal, and in scenario 1 this directly implies that the IF is not an accurate indicator of the value of an individual article.

The situation is very different in scenario 2. In this scenario, the articles in a journal all have a relatively similar value. The skewness of a journal citation distribution therefore does not result from large differences in the values of the articles in a journal. Instead, it results from the inaccuracy of citations as an indicator of the value of an article. As a consequence of this inaccuracy, articles that have a similar value may have very different numbers of citations. In line with the literature on cumulative advantage (De Solla Price, 1976) or preferential attachment (Barabási & Albert, 1999) processes, this causes the citation distribution of a journal to be skewed even though the articles in the journal all have a relatively similar value.

Like in scenario 1, in scenario 2 the IF is not representative of the number of citations of an individual article in a journal. However, this is not a problem in scenario 2. If a journal has published a sufficiently large number of articles, the IF may be expected to be a quite accurate indicator of the average value of the articles in the journal. This is the case despite the fact that in scenario 2 the number of citations of an individual article is a relatively inaccurate indicator of the value of the article. To understand this, it is essential to recognize that the IF is calculated at the level of an entire journal rather than at the individual article level. At the journal level, 'errors' in citations may be expected to largely cancel out. This is in agreement with what

Nicolaisen (2007) refers to as the standard account of citation analysis (e.g., Van Raan, 1998). If 'errors' in citations largely cancel out at the journal level, the IF is a quite accurate indicator of the average value of the articles in a journal. Since the articles in a journal all have a relatively similar value in scenario 2, this implies that the IF is also a quite accurate indicator of the value of an individual article.

### 4.2. Which scenario is more realistic?

Critics of the use of the IF for assessing individual articles implicitly appear to assume that reality is like scenario 1. Critics do not seem to be aware of the possibility of reality being more like scenario 2, or alternatively, they may consider scenario 2 to be highly unrealistic and may therefore not take it seriously. In our view, there is no easy way to determine whether scenario 1 or scenario 2 is closer to reality. Nevertheless, we can make some comments on the degree to which scenarios 1 and 2 may be realistic.

We first consider the accuracy of citations as an indicator of the value of an article. In scenario 1 citations are a relatively accurate indicator of the value of an article, while in scenario 2 they are a relatively inaccurate indicator. There are two reasons why it is difficult to say which of the two scenarios is more realistic.

First, there are conflicting viewpoints on the accuracy of citations as an indicator of the value of an article. For instance, following the well-known distinction between the normative and the social constructivist perspectives on citations (Nicolaisen, 2007), it is clear that those who adopt the normative perspective will have more confidence in the accuracy of citations than those who adopt the social constructivist perspective. Hence, followers of the normative perspective will be more likely to accept the viewpoint of scenario 1 on the accuracy of citations, while followers of the social constructivist perspective will reject this viewpoint and may find the viewpoint of scenario 2 more acceptable (although they may also disagree with this viewpoint).

There is a second reason why it is difficult to say which of the two scenarios provides a more realistic perspective on the accuracy of citations. As discussed in Subsection 3.1, we have chosen not to provide a precise definition of the concept of the value of an article. However, depending on how this concept is understood, one may prefer either scenario 1 or scenario 2. For instance, if the value of an article is understood as the extent to which the article is used in other articles, citations may perhaps be considered a relatively accurate indicator of the value of an article. From this point of view, scenario 1 may then be regarded as more realistic than scenario 2. On the other hand, if the value of an article is understood as the quality of the article according to the judgment of scientific peers, citations may be considered a relatively inaccurate indicator of the value of an article. Scenario 2 may then be regarded as more realistic than scenario 1.

We now consider the homogeneity or heterogeneity of journals in terms of the values of the articles they publish. In scenario 1 there are large differences in the values of the articles published in a journal, while in scenario 2 the articles published in a journal all have a relatively similar value.

The homogeneity of journals in scenario 2 can be motivated based on two ideas. One idea is that the peer review system of a journal will ensure that all or almost all articles in a journal have a value above a certain journal-specific minimum threshold. The other idea is that researchers will generally try to publish their work in a journal that is as 'prestigious' as possible, which means that they will try to avoid publishing their work in a journal that also publishes work of much lower value. Together, these two ideas may cause journals to be relatively homogeneous in terms of the values of the articles they publish.

The above motivation for the homogeneity of journals in scenario 2 requires a relatively high level of confidence in the accuracy of the journal peer review system. However, the accuracy of the journal peer review system has been questioned (for an overview of the literature, see Bornmann, 2011), which provides support for the heterogeneity of journals in scenario 1. There are also other arguments that may be used to support the heterogeneity of journals. For instance, when a journal publishes lots of articles, it seems unlikely that these articles are all of similar value. In general, the larger a journal, the more the journal can be expected to be heterogeneous in terms of the values of its articles. In addition, in a small field with only a limited number of journals (such as the field of scientometrics), even a relatively small journal may need to publish articles that are of quite different value. This also results in journals being heterogeneous.

We have now made a number of comments on the degree to which scenarios 1 and 2 may be realistic. Based on these comments, which of the two scenarios is closer to reality? In our opinion, there is no easy answer to this question. The answer is likely to be field- and journal-dependent. It is also likely to be time-dependent. In line with some of the literature discussed in Subsection 2.2 (Abramo et al., 2010; Ancaiani et al., 2015; Levitt & Thelwall, 2011), shortly after an article has been published, scenario 1 seems unrealistic, since there has not been much time for the article to be cited. Scenario 1 may be more realistic in the longer term. Scenario 2, on the other hand, may be realistic both in the short term and in the longer term. Furthermore, as we have already pointed out, the answer to the above question also depends on the precise understanding that one has of the concept of the value of an article. In other words, the appropriateness of the use of the IF for assessing individual articles is dependent on the precise criterion based on which one wants articles to be assessed.

Importantly, whether the IF can or cannot be used for assessing individual articles is perhaps not even the most relevant question to ask. Any method for assessing articles has weaknesses. This applies not only to the IF but also to the number of citations of an article and to assessment based on peer review. The question whether the use of a specific method for assessing articles is appropriate or not therefore seems to be of limited relevance. Instead, a more relevant question seems to be which of the various methods available for assessing articles is most appropriate *relative to the others*. For instance, critics of IF-based assessment of individual articles typically seem to believe that for assessing an article it is more appropriate to use the number of citations of the article than the IF of the journal in which the article has appeared. This seems to be the case for the critics quoted in Subsection 2.1, although some of them are more explicit about this than others. Gingras (2016, p. 48) is very explicit: "If one wants to measure the quality or visibility of a particular item, one must look at the citations actually received in the years following its publication." In our opinion, determining the relative appropriateness of different methods for assessing articles is a much more intricate problem than critics of IF-based assessment seem to believe. We reject a simple binary perspective in which some methods are valid and others are invalid. Instead, it is a matter of degree. Depending on the assumptions that one makes, one method may be more appropriate than another, but the difference need not be large. Also, the situation may reverse when the assumptions are changed. In the next section, we will use computer simulations to further elaborate our viewpoint.

## 5. Computer simulations

We now use computer simulations to further illustrate the ideas introduced in the previous two sections. We start by presenting our simulation model and by discussing how we analyze the accuracy of an indicator for assessing individual articles. We then report the results of our computer simulations.[2]

### 5.1. Model

We consider a scientific field in which there are $m$ journals. In a certain time period, $n$ articles are published in these journals. Each journal is of the same size, so each journal publishes $n/m$ articles.

For each article $i$ ($i = 1,2, \ldots, n$), the value of the article, denoted by $v_i$, is drawn from a lognormal distribution, that is,

$$v_i \sim \mathrm{logN}(\sigma_v^2). \qquad (1)$$

We use $\mathrm{logN}(\sigma^2)$ to denote a lognormal distribution for which the mean and the variance of the underlying normal distribution are equal to $-\sigma^2/2$ and $\sigma^2$, respectively. In this way, the mean of the lognormal distribution always equals 1, regardless of the value of $\sigma^2$. A lognormal distribution is used in (1) because in reality there are probably many more articles that have a low or moderate value than articles that have a high value. This is captured by the skewness of the lognormal distribution. The degree to which the distribution is skewed is determined by the parameter $\sigma_v^2$ in (1).

Journal 1 is regarded as the most prestigious journal in the field, journal 2 is regarded as the second most prestigious journal in the field, and so on. Journal $m$ is seen as the least prestigious journal. Our model does not specify why one journal is regarded as more prestigious than another journal. However, one could imagine that this is based on the IFs of the journals in earlier time periods or on the value of the articles published in the journals in earlier time periods. Our model assumes that the authors of an article first try to publish their article in journal 1. If their article is rejected by this journal, they try to publish it in journal 2, and so on. This goes on until there is a journal that accepts the article.

---

[2] Our use of computer simulations is somewhat related to work by Kapeller and Steinerberger (2016). Kapeller and Steinerberger use computer simulations to study the journal publishing system. Their focus is on analyzing the efficiency of the system, not on analyzing the accuracy of indicators for assessing individual articles.

To decide which articles to accept and which ones to reject, a journal $k$ estimates the value of each article it receives. To do so, the journal sends each article to reviewers. Based on the comments reviewers provide on an article, the journal obtains an estimate of the value of the article. The value of article $i$ estimated by journal $k$, denoted by $e_{ik}$, equals the value of the article multiplied by a value drawn from a lognormal distribution. More precisely, $e_{ik}$ is given by

$$e_{ik} = v_i \varepsilon_{ik} \text{ where } \varepsilon_{ik} \sim \text{logN}(\sigma_r^2). \quad (2)$$

The parameter $\sigma_r^2$ determines the accuracy of the journal peer review system. The smaller the value of this parameter, the more accurate the journal peer review system. If $\sigma_r^2 = 0$, the journal peer review system provides a perfectly accurate estimate of the value of an article. Of all articles received by journal $k$, the journal accepts the $n/m$ articles that have the highest estimated value. All other articles are rejected. Hence, journal 1 receives $n$ articles and rejects $n - n/m$ of them, journal 2 receives $n - n/m$ articles and rejects $n - 2(n/m)$ of them, and so on. Journal $m$, the least prestigious journal, receives $n/m$ articles, which it all accepts.

After all $n$ articles have been published, they accumulate citations. Our model assumes that the number of citations of an article correlates with the value of the article. On average, articles that have a higher value receive more citations. For each article $i$, the number of citations of the article, denoted by $c_i$, equals the value of the article multiplied by a value drawn from a lognormal distribution, that is,

$$c_i = v_i \varepsilon_i \text{ where } \varepsilon_i \sim \text{logN}(\sigma_c^2). \quad (3)$$

The parameter $\sigma_c^2$ determines the accuracy of citations as an indicator of the value of an article. The smaller the value of this parameter, the higher the accuracy of citations. If $\sigma_c^2 = 0$, citations are a perfectly accurate indicator of the value of an article. In reality, the number of citations of an article is an integer. For simplicity, we do not require the number of citations of an article to be an integer in our model. As we discuss in Subsection 5.3, citations may be interpreted as rescaled citations (Radicchi, Fortunato, & Castellano, 2008) in our model.

It follows from (1) and (3) that the distribution of citations over articles is also lognormal. More precisely, the distribution of citations over articles is $\log\mathrm{N}(\sigma_v^2 + \sigma_c^2)$. The lognormal distribution of citations over articles is in line with empirical studies that show that the distribution of citations over articles is highly skewed and approximately lognormal (Evans, Hopkins, & Kaube, 2012; Radicchi et al., 2008; Stringer, Sales-Pardo, & Amaral, 2008; Thelwall, 2016a, 2016b).

Finally, for each journal $k$, the IF of the journal, denoted by $IF_k$, is calculated. In our model, the IF of a journal is defined as the average number of citations of the articles published in the journal. Hence, $IF_k$ is given by

$$IF_k = \frac{\sum_{i=1}^{n} p_{ik} c_i}{\sum_{i=1}^{n} p_{ik}}, \tag{4}$$

where $p_{ik}$ equals 1 if article $i$ has been published in journal $k$ and 0 otherwise. In our model, each journal publishes $n/m$ articles, and therefore the denominator in (4) always equals $n/m$.

The model introduced above of course provides a simplified representation of reality. For instance, in reality journals are not all of the same size and researchers do not all have the same perception of the prestige of the journals in their field. Also, when researchers want to publish an article, they do not always start by submitting their article to the most prestigious journal. Based on their knowledge of the journals in their field, researchers may know the journal in which their article can best be published, and they may immediately submit their article to this journal rather than first submitting it to other more prestigious journals.

Importantly, we consider the simplicity of our model to be a strength, not a weakness. We could develop a more realistic model. However, such a model would also be more complex, making it more difficult to obtain clear insights from the model. A good model captures the essential elements that need to be taken into account to get a proper understanding of the phenomenon of interest, while it leaves out the non-essential elements. We believe that our model indeed captures the elements that are essential for our analysis. At the same time, non-essential elements are left out, so that unnecessary complexity is avoided and clear insights can be obtained.

### 5.2. Accuracy of an indicator

We focus on two indicators for assessing individual articles. One indicator is the IF of the journal in which an article has appeared.[3] The other indicator is the number of citations of an article. Our aim is to analyze and compare the accuracy of these two indicators. This of course requires a precise definition of the accuracy of an indicator.

Our definition of the accuracy of an indicator relies on a binary classification of articles based on their value. Like in Subsection 3.2, we distinguish between low-value and high-value articles. To make this distinction, we introduce the parameter $\alpha$. This parameter specifies the share of articles that are considered to be of high value. Of the $n$ articles in our simulation model, the $\alpha n$ articles with the highest values are classified as high-value articles, while the remaining articles are classified as low-value articles.

To obtain the accuracy of an indicator, we select the $\alpha n$ articles that are most highly ranked by the indicator (i.e., the $\alpha n$ articles with the highest IF or the largest number of citations) and we calculate the percentage of the selected articles that are of high value. The accuracy of an indicator can be anywhere between $0\%$ and $100\%$. An indicator has an accuracy of $100\%$ if the $\alpha n$ articles that are most highly ranked by the indicator coincide with the $\alpha n$ high-value articles. An indicator has an accuracy of $0\%$ if the $\alpha n$ most highly ranked articles are all of low value.

### 5.3. Results

We now present the results of our computer simulations. We consider a situation in which $n = 2000$ articles are published in a certain scientific field and in a certain time period. These articles appear in $m = 20$ journals, which means that each journal publishes $n/m = 2000/20 = 100$ articles.

---

[3] In reality, the way in which the IF is used for assessing individual articles is slightly different from the way in which this is done in our simulation model. In reality, when an article published in year $y$ is assessed using the IF, the IF is calculated based on citations received by articles published in the same journal in years $y - 1$ and $y - 2$. To keep our simulation model as simple as possible, time is not taken into account in the model. Essentially, in our model, the IF is calculated based on citations received by articles published in year $y$ rather than in years $y - 1$ and $y - 2$. Although our model provides a simplified representation of reality, this does not affect our analysis in an essential way. The key element in the discussion on the use of the IF for assessing individual articles is the skewness of citation distributions, and this skewness is properly reproduced in our model.

To choose suitable values of $\sigma_v^2$ and $\sigma_c^2$, we rely on empirical work carried out by Radicchi et al. (2008). Radicchi et al. rescale citations in such a way that in each field the average number of citations per article equals 1. This is in agreement with our simulations, in which we also have an average number of citations per article of 1. Radicchi et al. report that the distribution of rescaled citations over the articles in a field is lognormal, with the variance of the underlying normal distribution being equal to 1.3.[4] In order to obtain citation distributions that are in line with the findings of Radicchi et al., we require that $\sigma_v^2 + \sigma_c^2 = 1.3$ in our simulations. In the presentation of the simulation results, we report the value of $\sigma_c^2$. Values of $\sigma_c^2$ between 0 and 1.3 are considered. The value of $\sigma_v^2$ is not reported, but this value equals $1.3 - \sigma_c^2$. Suitable values of $\sigma_r^2$ cannot be easily derived from empirical analyses. We therefore simply consider a number of different values of $\sigma_r^2$ in our simulations.

In the calculation of the accuracy of an indicator, we set the parameter $\alpha$ equal to 0.1. Hence, we determine the accuracy of an indicator based on the capability of the indicator to identify the 10% highest-value articles. The choice of $\alpha = 0.1$ is somewhat arbitrary. However, we also tested other values of $\alpha$, and our results do not change in an essential way when a different value of $\alpha$ is chosen. Our simulation results are based on 1000 simulation runs. The accuracy of an indicator is calculated as the average accuracy over all simulation runs.

Figure 1 shows for different values of $\sigma_r^2$ and $\sigma_c^2$ the accuracy of both the IF of the journal in which an article has appeared and the number of citations of an article. Four different values of $\sigma_r^2$ are considered. In our simulation model, the value of $\sigma_r^2$ has no influence on the accuracy of citations, but it does influence the accuracy of the IF. The higher the value of $\sigma_r^2$ (i.e., the lower the accuracy of the journal peer review system), the less accurate the IF. As can be expected, the accuracy of both the IF and citations decreases as the value of $\sigma_c^2$ increases. However, the value of $\sigma_c^2$ has more influence on the accuracy of citations than on the accuracy of the IF. As discussed in

[4] The findings of Radicchi et al. are criticized by Waltman, Van Eck, and Van Raan (2012), who show that rescaled citation distributions do not have exactly the same shape in different fields. Nevertheless, the findings of Radicchi et al. provide a reasonable approximation of the true shape of citation distributions, and we therefore use these findings to inform the choice of the values of $\sigma_v^2$ and $\sigma_c^2$ in our simulations.

Subsection 4.1, this is because ‘errors’ in citations tend to cancel out in the IF, making the IF relatively insensitive to these ‘errors’.

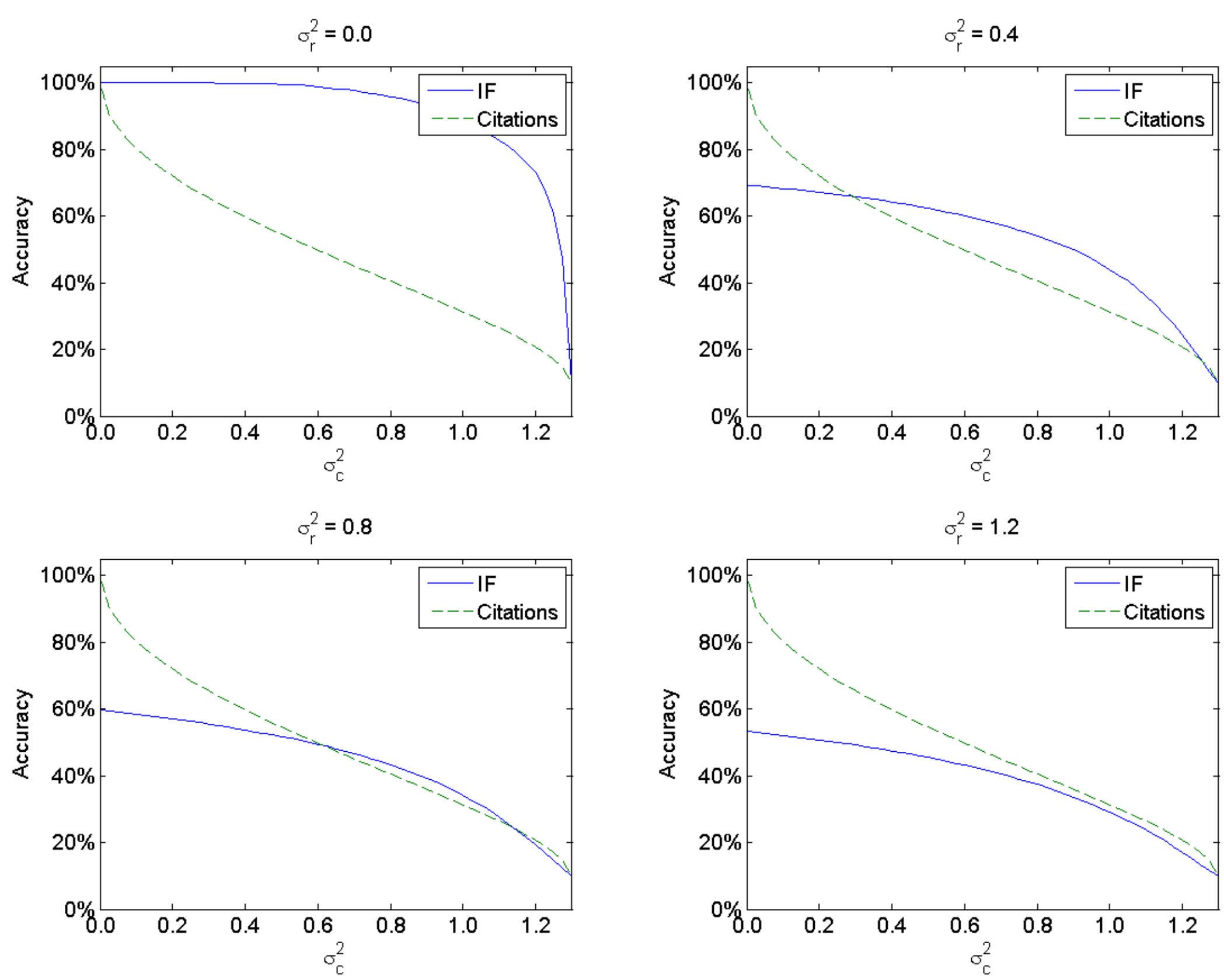

Figure 1. Accuracy of the IF and of citations for different values of $\sigma_r^2$ and $\sigma_c^2$.

The most important observation based on Figure 1 is that for a range of values of $\sigma_r^2$ and $\sigma_c^2$ the IF is more accurate than citations. This is the case when the value of $\sigma_r^2$ is not too high (i.e., the journal peer review system is at least moderately accurate) and the value of $\sigma_c^2$ is not too low (i.e., citations are at least moderately inaccurate). For these values of $\sigma_r^2$ and $\sigma_c^2$, the IF benefits from its limited sensitivity to ‘errors’ in citations while it does not suffer too much from heterogeneity in the values of the articles published in a journal. As shown in the top-left panel in Figure 1, when $\sigma_r^2 = 0$ (i.e., the journal peer review system is perfectly accurate), the IF outperforms citations for all values of $\sigma_c^2$. On the other hand, the bottom-right panel in Figure 1 shows that for high values of $\sigma_r^2$ (i.e., the journal peer review system is highly inaccurate) the IF is always outperformed by citations, regardless of the value of $\sigma_c^2$. In this case, journals are highly heterogeneous and publish a mix of high-value and low-value articles, making the IF a very weak indicator of the value of an article.

The results presented in Figure 1 are based on a situation in which there are $m = 20$ journals. Figure 2 shows the effect of increasing or decreasing the number of journals, while keeping the total number of articles fixed at $n = 2000$. In the left panel of Figure 2, the number of journals has been halved (and the number of articles per journal has been doubled), which means that we have $m = 10$ journals with $n/m = 200$ articles per journal. In the right panel, the number of journals has been doubled (and the number of articles per journal has been halved), resulting in $m = 40$ journals with $n/m = 50$ articles per journal. In both panels, $\sigma_r^2$ has a value of 0.4. Hence, we consider an intermediate level of accuracy of the journal peer review system.

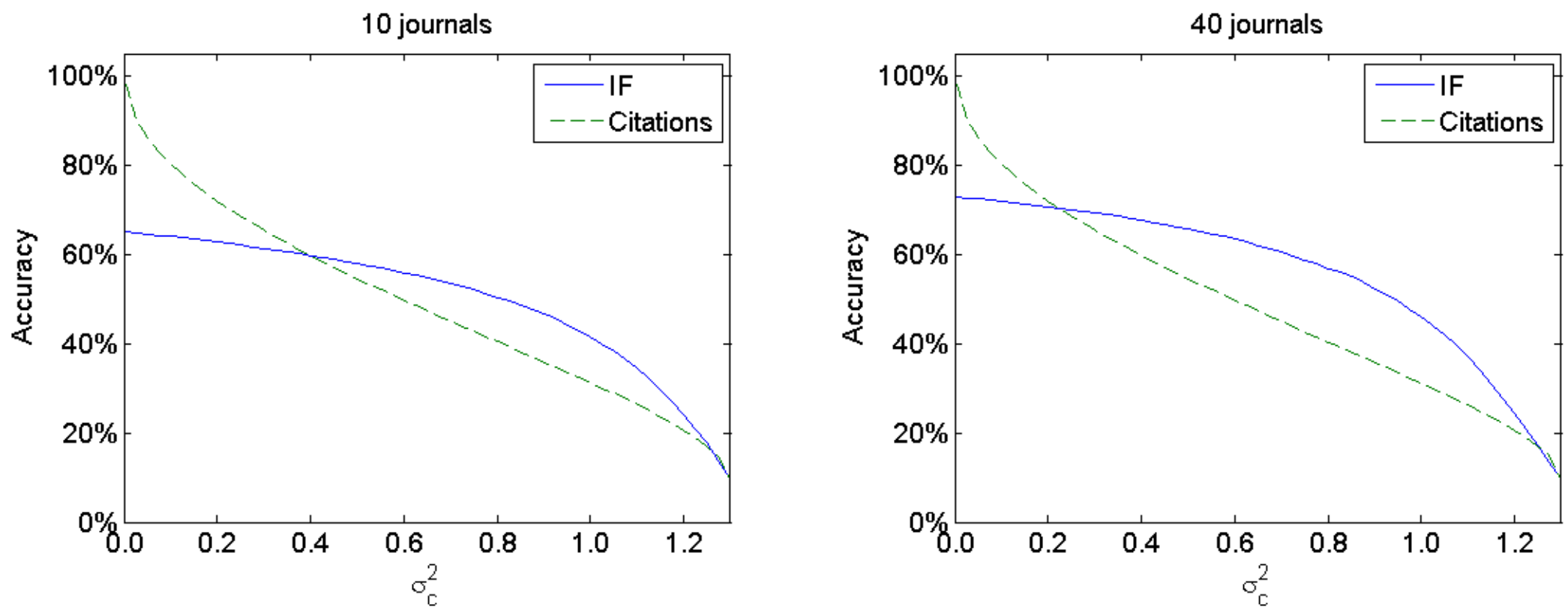

Figure 2. Accuracy of the IF and of citations for different numbers of journals and for different values of $\sigma_c^2$.

Increasing the number of journals from 10 (left panel of Figure 2), to 20 (top-right panel of Figure 1), to 40 (right panel of Figure 2) yields a modest improvement in the accuracy of the IF. Of course, it does not affect the accuracy of citations. The increase in the number of journals therefore broadens the range of values of $\sigma_c^2$ in which the IF outperforms citations. When the IF is used as an indicator of the value of an article, it is clear that the number of journals should not be too small. In the extreme case in which there is only one journal (i.e., $m = 1$), the IF is completely useless as an indicator of the value of an article. However, the number of journals should not be too large either. Having a large number of journals is fine as long as the number of articles per journal does not become too small. When the number of articles per journal is very small, the IF will be highly sensitive to 'errors' in citations. The smaller the number of articles in a journal, the less one can expect 'errors' in citations to cancel out. In the extreme case in which each journal publishes only one article (i.e., $m = n$), the IF and citations have exactly the same accuracy. However, this is an artefact that results from our choice not to take time into account in our simulation model (see footnote 3).

We have seen that, depending on the values of $\sigma_r^2$, $\sigma_c^2$, and $m$, the accuracy of the IF may be either higher or lower than the accuracy of citations. A natural question to ask is whether the IF and citations can be combined into a hybrid indicator that is more accurate than both the IF and citations separately. This possibility was already suggested by Levitt and Thelwall (2011) and Anfossi, Ciolfi, Costa, Parisi, and Benedetto (2016). To explore this possibility, we obtain hybrid indicators by calculating a weighted average of the IF of the journal in which an article has appeared and the number of citations of the article.[5] We give a weight of $0\%$, $25\%$, $50\%$, $75\%$, or $100\%$ to the IF. The remaining weight is given to citations. Of course, when the IF has a weight of $0\%$, the hybrid indicator coincides with the citations indicator. Likewise, using a weight of $100\%$ for the IF, the hybrid indicator coincides with the IF indicator. We focus on the situation in which $\sigma_r^2 = 0.4$ and $m = 20$.

The results are presented in Figure 3. The figure confirms that hybrid indicators indeed perform well. Except for very low values of $\sigma_c^2$, citations are consistently outperformed by a hybrid indicator that gives a weight of $25\%$ to the IF and a weight of $75\%$ to citations. The other way around, for any value of $\sigma_c^2$, the IF is outperformed by a hybrid indicator that gives a weight of $75\%$ to the IF and a weight of $25\%$ to citations. These results show that one does not necessarily need to make an absolute choice between the IF and citations. Instead, the two indicators can be combined into a hybrid indicator that is likely to be more accurate than each of the two indicators separately.

[5] In our simulation model, the IF and citations have the same scale (i.e., they both have an average value of 1) and therefore it makes sense to combine them in a straightforward way by calculating a weighted average. In practice, the IF and citations are likely to have different scales. This needs to be accounted for when combining them into a hybrid indicator.

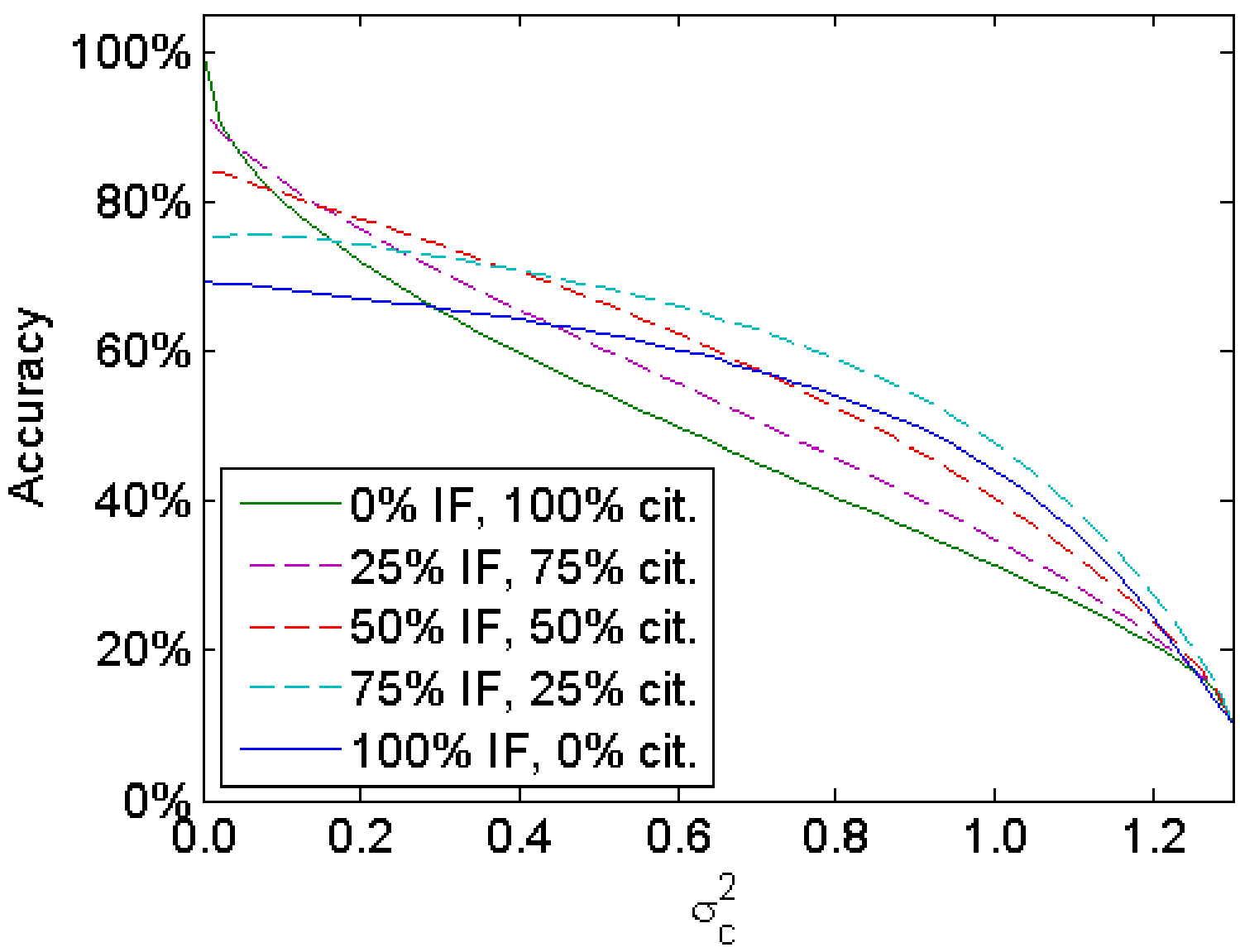

Figure 3. Accuracy of different hybrid indicators combining the IF and citations for different values of $\sigma_c^2$.

## 6. Discussion and conclusion

According to Van Raan (quoted by Van Noorden, 2010, p. 864–865), "if there is one thing every bibliometrician agrees, it is that you should never use the journal impact factor to evaluate research performance for an article or for an individual — that is a mortal sin". As discussed in Section 2, many scientometricians indeed reject the use of the IF for assessing individual articles. Moreover, the widespread support for DORA (2013) shows that the same applies to the scientific community more generally. Many objections against the IF and the way in which it is used in research evaluations are legitimate and deserve careful consideration. However, the statistical arguments that are typically employed to reject the use of the IF at the level of individual articles are not convincing. A more nuanced perspective on the use of the IF and of journal-level indicators more generally is therefore needed.

As we have shown using an illustrative example in Section 3, a conceptual discussion in Section 4, and computer simulations in Section 5, commonly used statistical arguments against the use of the IF for assessing individual articles are incorrect. This applies to arguments based on the skewness of citation distributions, and it also applies to other related types of arguments, such as the ecological fallacy argument (Leydesdorff et al., 2016; Paulus et al., 2018). Although these arguments may appear compelling at first sight, a more careful analysis reveals that the arguments do not logically lead to the conclusion that the IF should not be used at the level of individual articles.[6] These arguments lead to this conclusion only when additional assumptions are made, for instance the assumption that citations accurately reflect the value of an article or the assumption that journals are very heterogeneous in terms of the values of the articles they publish. Our analysis not only shows that statistical objections against the use of the IF at the level of individual articles are not convincing. It also shows that, depending on the assumptions that are made, the IF may be a more accurate indicator of the value of an article than the number of citations of the article. We emphasize that our analysis does not address other concerns one may have about the IF, for instance related to its calculation or transparency.

Our analysis is of a theoretical nature, and it therefore does not make clear whether *in practice* the use of the IF for assessing individual articles is to be recommended or not and whether *in practice* the IF is more or less accurate than citations. These questions require empirical follow-up research. One could for instance compare the accuracy of the IF and of citations by correlating both of them with peer review assessments of articles. Such an analysis is presented by HEFCE (2015) (for similar analyses at a smaller scale, see Allen, Jones, Dolby, Lynn, & Walport, 2009; Eyre-Walker & Stoletzki, 2013). The analysis is based on the outcomes of the Research Excellence Framework in the United Kingdom. It shows that two field-normalized journal-level indicators, SNIP and SJR, and field-normalized citations all correlate more or less to the same degree with peer review assessments. However, in this analysis, peer review took place after articles had been published, and therefore peer review assessments may have been influenced by the fact that reviewers knew in which journal an article had appeared and how often an article had been cited. Ideally, when using peer review assessments to compare the accuracy of the IF and of citations, one would like the peer review assessments to be completely independent of this type of information.

Follow-up research may also focus on developing more advanced simulation models for analyzing the use of the IF in research evaluations. The model presented in

[6] We refer to Schauer (2003) for a related discussion from a moral philosophical perspective. Schauer points out that 'particularism' (e.g., assessing an article based on the number of citations it has received) is not necessarily preferable over 'generalization' (e.g., assessing an article based on the IF of the journal in which it has appeared).

Section 5 is static and involves only a single time period. In a dynamic model with multiple time periods, the IF of a journal can be calculated in a more realistic way (by using appropriate publication and citation windows) and may evolve over time. Moreover, in a dynamic model, the citations of the articles published in a journal may not only determine the IF of the journal but may also be influenced by the IF of the journal in earlier time periods, creating a kind of Matthew effect of the IF (for further discussion on this possibility, see Kim, Portenoy, West, & Stovel, 2020; Larivière & Gingras, 2010; Traag, in press). A more advanced simulation model may also consider that the peer review carried out by journals takes time and that researchers may not want to risk delaying publication of their work by submitting it to a journal by which it will most likely be rejected. Hence, researchers may make their own assessment of the value of their work, and based on this they may choose a suitable journal to which they submit their work. Another idea that can be considered in a more advanced simulation model is that even within a single field of science journals may differ significantly in their topical focus. This influences how researchers choose the journal to which they submit their work. The situation becomes especially complex when some topics attract more citations than others. The IF may then create an incentive both for journals and for researchers to shift their attention to specific topics (e.g., Müller & De Rijcke, 2017). A final possibility for a more advanced simulation model is to regard the IF and citations as proxies of different aspects of the value of an article, leading to a situation in which the IF and citations may be seen as two complementary indicators that each provide useful information.

Importantly, the simplicity of the simulation model presented in Section 5 does not weaken our claim that commonly used statistical arguments against the use of the IF for assessing individual articles are incorrect. We received a lot of feedback on an earlier version of this paper, which we published as a preprint in 2017 (Waltman & Traag, 2017). Many of the critical comments that we received were about unrealistic assumptions in our simulation model. We agree that the model is unrealistic in many ways, and we are open to the possibility that more realistic models may provide stronger arguments against the use of the IF at the level of individual articles. However, this does not invalidate our claim that commonly used statistical arguments against the use of the IF at the article level are incorrect. Our simulation model captures the essence of these arguments and shows that they are not convincing.

In our view, whether article-level use of the IF is justified or not cannot be decided based only on statistical arguments. For instance, in our simulation model, the appropriateness of article-level use of the IF depends on the assumptions that one makes, and whether certain assumptions are realistic or not is an empirical rather than a statistical matter. This illustrates that arguments against article-level use of the IF should not be based exclusively on statistical considerations. These arguments need to be supported by careful empirical analyses. While some of the statistical properties of the IF may be questioned, we do not agree with the idea that the use of the IF at the article level reflects 'statistical illiteracy' (Curry, 2012).

Based on their study of the use of the IF in university medical centers in the Netherlands, Rushforth and De Rijcke (2015) state that they "feel ambivalent about statements coming from scientometricians that the JIF 'misleads.' By limiting indicator uses to questions of validity, movements like DORA also assume displacing the JIF for 'better' (i.e. more valid) indicators would necessarily give rise to better evaluation practices" (p. 136). In a similar spirit, Cronin and Sugimoto (2015) consider the use of the IF for assessing individual articles to be "as much a socio-technical as a statistical issue: growing adoption of the IF is changing scientists' behavior and causing displacement activity". While the use of the IF for assessing individual articles need not be statistically wrong, it often seems highly problematic from a socio-technical perspective. We have not considered the socio-technical perspective in this paper, but we recognize that there is ample evidence of undesirable consequences of the dominant role of the IF in research evaluations (e.g., Chorus & Waltman, 2016; Larivière & Sugimoto, 2019; Martin, 2016; Wilhite & Fong, 2012).

We are in full support of initiatives aimed at improving research evaluation, and we believe that critical discussions about the use of the IF should be an important element of such initiatives. However, it is not clear whether replacing the IF by article-level indicators will improve research evaluation. If article-level indicators become as dominant as the IF, this can be expected to have undesirable consequences similar to those of the pervasive use of the IF.[7] The use of the IF and other journal-

[7] In addition, some article-level indicators also have the disadvantage of being less transparent than the IF. For instance, as discussed by Steck, Stalder, and Egger (2020), the Medical Faculty of the University of Bern has signed DORA and, as a result, has replaced the IF by an article-level indicator, the relative citation ratio (RCR; Hutchins, Yuan, Anderson, & Santangelo, 2016). In our view, it is

level indicators (Wouters et al., 2019) in research evaluations needs to be critically discussed, but the discussion should not be based on misplaced statistical arguments.

## Acknowledgements

We have received feedback on our work from a large number of colleagues. In addition to our colleagues at the Centre for Science and Technology Studies at Leiden University, this includes Giovanni Abramo, Sergio Benedetto, Lutz Bornmann, Kevin Boyack, Stephen Curry, Phil Davis, Adam Eyre-Walker, Jochen Gläser, Robin Haunschild, Michael Kurtz, Vincent Larivière, Loet Leydesdorff, Charles Oppenheim, Andrew Plume, Ismael Rafols, Ronald Rousseau, Javier Ruiz-Castillo, Jesper Schneider, Cassidy Sugimoto, and Mike Thelwall. We are grateful for all feedback. This has led to important improvements to our work.

---

questionable whether this is an improvement. The RCR is less transparent than the IF and has received significant criticism (Janssens, Goodman, Powell, & Gwinn, 2017; Waltman, 2015).